\documentclass[fleqn,11pt,twoside]{article}

\usepackage{amsthm,amsthm,amssymb, color, xcolor,epsfig, graphics, subfigure}

\usepackage{amsmath, graphicx, latexsym, lscape }

\usepackage{cite}

\makeatletter
\newcommand{\copyrightnote}[2]{{\renewcommand{\thefootnote}{}
 \footnotetext{\small\it
\begin{flushleft}
 \copyright \ #1   #2
\end{flushleft}}}}

\newcommand{\Name}[1]{\begin{flushleft}
                       \LARGE \bf #1
                       \end{flushleft}\vspace{-3mm}}

\newcommand{\Author}[1]{\begin{flushleft}
                       \it #1 \end{flushleft}}

\newcommand{\Address}[1]{\begin{flushleft}
                       \it #1 \end{flushleft}}

\newcommand{\Date}[1]{\begin{flushleft}
                      \small  \it #1 \end{flushleft}}

\newcommand{\evenhead}{Author \ name}
\newcommand{\oddhead}{Article \ name}

\renewcommand{\@evenhead}{
\hspace*{-3pt}\raisebox{-15pt}[\headheight][0pt]{\vbox{\hbox to \textwidth
{\thepage \hfil \evenhead}\vskip4pt \hrule}}}
\renewcommand{\@oddhead}{
\hspace*{-3pt}\raisebox{-15pt}[\headheight][0pt]{\vbox{\hbox to \textwidth
{\oddhead \hfil \thepage}\vskip4pt\hrule}}}
\renewcommand{\@evenfoot}{}
\renewcommand{\@oddfoot}{}

\long\def\@makecaption#1#2{%
  \vskip\abovecaptionskip
  \sbox\@tempboxa{\small \textbf{#1.}\ \ #2}%
  \ifdim \wd\@tempboxa >\hsize
    {\small \textbf{#1.}\ \ #2}\par
  \else
    \global \@minipagefalse
    \hb@xt@\hsize{\hfil\box\@tempboxa\hfil}%
  \fi
  \vskip\belowcaptionskip}

\newcommand{\JNMPnumberwithin}[3][\arabic]{%
  \@ifundefined{c@#2}{\@nocounterr{#2}}{%
    \@ifundefined{c@#3}{\@nocnterr{#3}}{%
      \@addtoreset{#2}{#3}%
      \@xp\xdef\csname the#2\endcsname{%
        \@xp\@nx\csname the#3\endcsname .\@nx#1{#2}}}}%
}

\newcommand{\resetfootnoterule} {
  \renewcommand\footnoterule{%
  \kern-3\p@
  \hrule\@width.4\columnwidth
  \kern2.6\p@}
}

\renewcommand{\footnoterule}{}

\makeatother

\theoremstyle{definition}

\usepackage{color}
\usepackage{cite}
\usepackage{makeidx}

\def\a{\alpha}
\def\b{\beta}
\def\ga{\gamma}

\def\vphi{\varphi}
\def\la{\lambda}

\def\s{\sigma}

\def\vphi{\varphi}

\def\De{\Delta}

\def\pa{\partial}

\def\o+{\oplus}

\def\<{\langle}
\def\>{\rangle}

\def\({\left(}
\def\){\right)}
\def\[{\left[}
\def\]{\right]}
\def\=#1{\bar #1}
\def\~#1{\widetilde #1}

\def\.#1{\dot #1}
\def\^#1{\widehat #1}

\def\"#1{\ddot #1}

\def\eeq{\end{equation}}
\def\beq{\begin{equation}}

\def\beql#1{\begin{equation} \label{#1}}

\def\eqref#1{(\ref{#1})}

\def\symmref{AVL,CGbook,KrV,Olver1,Olver2,Stephani}
\def\sderef{Arnold,Evans,Fre,Ikeda,Kampen,Oksendal,Stroock,Amir,Gardiner}

\def\FP{Fokker-Planck }

\begin{document}

\renewcommand{\evenhead}{ {\LARGE\textcolor{blue!10!black!40!green}{{\sf \ \ \ ]ocnmp[}}}\strut\hfill G. Gaeta \& M.A. Rodr\'iguez}
\renewcommand{\oddhead}{ {\LARGE\textcolor{blue!10!black!40!green}{{\sf ]ocnmp[}}}\ \ \ \ \ Integrable Ito and associated Fokker-Planck equations}

\thispagestyle{empty}
\newcommand{\FistPageHead}[3]{
\begin{flushleft}
\raisebox{8mm}[0pt][0pt]
{\footnotesize \sf
\parbox{150mm}{{Open Communications in Nonlinear Mathematical Physics}\ \  \ {\LARGE\textcolor{blue!10!black!40!green}{]ocnmp[}}
\ \ Vol.3 (2023) pp
#2\hfill {\sc #3}}}\vspace{-13mm}
\end{flushleft}}

\FistPageHead{1}{\pageref{firstpage}--\pageref{lastpage}}{ \ \ Article}

\strut\hfill

\strut\hfill

\copyrightnote{The author(s). Distributed under a Creative Commons Attribution 4.0 International License}

\Name{Integrable Ito equations and properties of the associated Fokker-Planck equations}

\Author{G. Gaeta$^{\,1}$ and M.A. Rodr\'iguez$^{\,2}$}

\Address{$^{1}$ Dipartimento di Matematica, Universit\`a degli Studi di
Milano, via Saldini 50, 20133 Milano (Italy)  {\it and} SMRI, 00058 Santa Marinella (Italy); {\tt giuseppe.gaeta@unimi.it} \\[2mm]
$^{2}$ Departamento de F\'{\i}sica Te\'orica,
Universidad Complutense de Madrid, Pza.~de las Ciencias 1, 28040 Madrid (Spain); {\tt rodrigue@ucm.es} }

\Date{Received May 26, 2023; Accepted June 16, 2023}

\setcounter{equation}{0}

\begin{abstract}

\noindent In a recent paper we have classified scalar Ito equations which admits a standard symmetry; these are also directly integrable by the Kozlov substitution. In the present work, we consider the diffusion (Fokker-Planck) equations associated to such symmetric Ito equations.

\end{abstract}

\label{firstpage}


\section{Introduction}

Symmetry methods \cite{\symmref} are a classical tool to attack nonlinear differential equations since the work of S. Lie. This statement, strictly speaking, refers to \emph{deterministic} differential equations; but in more recent years it also extended to \emph{stochastic} differential equations \cite{\sderef}.

In particular, work by R. Kozlov \cite{Koz1,Koz2,Koz3,Koz2018} showed that an Ito equation possessing a Lie-point symmetry of a suitable type (see below for details) is integrable; moreover, this result is constructive, in that once the symmetry has been determined one also knows the change of variables taking the equation into directly integrable form.

Classification of symmetry properties for Ito equations -- in particular, for the scalar Ito equations we are considering in this work -- has been performed several times, corresponding to considering different classes of ``admitted'' Lie-point symmetries; see, in particular, early work by Kozlov \cite{Koz2} considering a general class of what are now called \emph{deterministic} symmetries but allowing also transformations of time -- which are not allowed by the integration scheme which was then devised by Kozlov himself (his classification work had a more general scope).

More recently, it was observed that one can also consider so called \emph{random} symmetries \cite{GS17,Koz18a}; in particular integration of an Ito equation is guaranteed -- via the Kozlov scheme -- once a random or deterministic standard symmetry for this is known (the word ``standard'' is here meant as opposed to so called ``W-symmetries'', see below).

In a recent paper, we have classified all the scalar Ito equations admitting a (deterministic or random) standard symmetry, hence all the integrable scalar Ito equations \cite{GR22a}. This classification uses a well known preliminary  standard change of variables \cite{Oksendal} taking the noise coefficient of the Ito equation to unity; the classification is completely explicit in the case of autonomous equation and (in order to avoid exceedingly complicate formulas) only partially explicit in the case of time-dependent Ito equations. In the present paper -- albeit also considering the general case where appropriate -- we will first and mainly focus on autonomous  equations; this both for the sake of simplicity and because this suffices to make our main point.

As well known, the statistical properties of an Ito equation are described by the associated \emph{\FP equation}; this is also known as the \emph{Kolmogorov forward equation} \cite{Kampen,Gardiner,Risken}.

Given that \emph{integrable} Ito equations are characterized by having non-trivial symmetry properties, see above, it is natural to wonder if the corresponding \FP equations are also enjoying special symmetry properties. This
will be made more precise in the following, after describing what can be the symmetries of a general \FP equation;  but roughly speaking one could expect for the \FP equations associated to integrable Ito equations to find either a maximal Lie symmetry algebra, isomorphic to that of the heat equation (associated to the free Ito equation $dx = dw$); or at least a non-generic one (as the \FP equation is linear, certain symmetries will always be present, see below).

The second option, i.e. that the \FP equations associated to integrable Ito equations have a nontrivial Lie symmetry algebra, is actually seemingly implied by an old result by one of us and Rodr\'iguez-Quintero \cite{GRQ1,GRQ2}, stating that symmetries of an Ito equation (within a certain class, which includes what nowadays corresponds to deterministic symmetries, see below) are reflected into (projectable) symmetries of the associated \FP equation. Thus, given that one knows that generic \FP equations have a trivial symmetry algebra,
this weaker conjecture appears to be surely, and trivially, true; but, one should remember this is the case when we speak of the type of symmetries considered in that work, i.e. (in nowadays language) \emph{deterministic} symmetries.

It should be noted that not only the description of a system provided by these two languages, i.e. by Ito and by \FP equations, are intrinsically different from a physical standpoint, but the notions of integrability for an Ito equation and for the associated \FP equation are also substantially different.

In fact, integrability of an Ito equation means this is integrable for \emph{each} realization of the driving Wiener process; note that this does not imply that we can predict the behavior of the solution of an integrable Ito equation if we know the initial state, as we cannot predict the behavior of the driving Wiener process.

On the other hand, the description provided by the \FP equation does not consider specific realizations of the driving Wiener process but only the evolution of a general probability measure (thus an average -- in probabilistic sense -- over the possible realizations of the Wiener process), and integrability means that if the initial probability distribution is known, we can compute the probability distribution at any later time.

Thus, it is not \emph{apriori} obvious that there should be a correspondence between integrability, and \emph{a fortiori} symmetry,  properties of an Ito and of the associated \FP equations.

However one could, as a naive guess, expect that to an integrable Ito equation (such as the free one, whose solutions are just the realizations of Wiener process itself - and for which the associated \FP equation is just the heat equation) should correspond a \FP equation with maximal symmetry algebra (such as the heat equation). We will see this is not the case. Actually, it turns out that to an Ito equation which is integrable thanks to a \emph{random} symmetry may correspond a \FP equation which has a trivial symmetry.\footnote{One should recall, in this respect, that integrability through a random symmetry is properly speaking only formal, in that the change of variable taking the Ito equation to its explicitly integrable form actually maps it into a \emph{generalized} Ito equation. See the discussion in \cite{Gae23} for details.}

\section{General notations and background}
\label{sec:general}

We now first introduce general notation and then briefly describe background results on Ito equations, \FP
equations, and their symmetries.

\subsection{Ito equations}
\label{sec:gen:ito}

In this paper we consider \emph{scalar} Ito equations
\cite{\sderef}, routinely written as \beql{eq:Itogen} d x \ = \
f(x,t) \, dt \ + \ \s (x,t) \, d w \ ; \eeq here $w = w(t)$ is the
driving Wiener process, $f$ is the drift coefficient and $\s$ the
noise coefficient.\footnote{Note we might as well consider scalar equations depending on several independent driving Wiener processes, with functionally independent noise coefficients. This setting is considered in \cite{GR22b}; but will not be considered here.}

If $\s = 0$ the equation is actually a deterministic one, and we consider it to be a trivial Ito equation; we will only consider nontrivial Ito equations, i.e. always assume $\s \not= 0$.

We are specially interested in \emph{autonomous} scalar equations,
i.e. in the case where both $f$ and $\s$ depend on $x$ alone (possibly
being actually constant),
\beql{eq:Ito} dx \ = \ f (x) \, dt \ + \ \s (x) \, d w \ . \eeq
The present discussion of general features will be conducted first in the general case \eqref{eq:Itogen}, but then specialized to the autonomous case \eqref{eq:Ito}.

Integrable scalar Ito equations have been recently classified
\cite{GR22a}; as mentioned above a previous classification --
considering also time reparametrization, but not considering
random symmetries (see below) -- had been obtained by Kozlov \cite{Koz2}.

When looking for Lie-point symmetries of \eqref{eq:Itogen} or \eqref{eq:Ito}, we
consider vector fields of the form \beql{eq:Xgen} X \ = \ \vartheta (t)
\, \pa_t \ + \ \vphi (x,t;w) \, \pa_x \ . \eeq The functional form
of $\vartheta$ means that we admit at most a reparametrization of
time (see e.g. \cite{GS17,Koz18a,Koz18b,GSW,Gae23,GR22b,GRQ1,GRQ2,GR22a,GGPR,GL1,GL2} for a discussion of admissible transformations in this context); on the other hand,
admitting a $w$-dependence of $\vphi$ means that the transformation of the dynamical variable $x$ can depend not
only on $x=x(t)$ itself and on time $t$, but also on the value
$w(t)$ reached by the driving Wiener process at time $t$.

When $\vartheta = 0$ we have \emph{time-preserving symmetries}; they are related to the integrability of the Ito equation via the Kozlov substitution, see below, and hence we will focus on these.

For such symmetries, i.e. for symmetries generated by vector fields of the form \beql{eq:X}
X \ = \ \vphi (x,t;w) \ \pa_x \ , \eeq a generally accepted
nomenclature -- and anyway the one we will follow here -- is that: \begin{itemize}
    \item[(a)] when $\vphi_w = 0$ we have a \emph{deterministic
standard symmetry}, or simply a \emph{deterministic symmetry};
\item[(b)] when $\vphi_w \not= 0$ we have a \emph{random standard symmetry},
or simply a \emph{random symmetry}.\footnote{The qualification of
``standard'' symmetry serves not only to specify they are time-preserving but also to avoid confusion with so called W-symmetries \cite{GSW}; however we will not consider these in the present setting, as they are on the one hand not useful to integrate Ito equations (see the discussion in \cite{Gae23}), and on the other hand involve transformations of the Wiener process $w(t)$, which itself does not appear in the \FP equation. We will thus mostly omit, in the present work, the specification ``standard'' when referring to deterministic or
random symmetries.}
\end{itemize}

Symmetries are determined as solutions to the
\emph{determining equations}; these are a set of linear PDEs for $\vphi =
\vphi (x,t;w)$; in the scalar case (and for standard symmetries), these are explicitly written as
\begin{eqnarray}
\vphi_t & + & f \, \vphi_x \ - \ \vphi \, f_x \ + \ \frac12 \, \De
(\vphi ) \ = \ \vartheta \, f_t \ , \\
\vphi_w & + & \s \, \vphi_x \ - \ \vphi \, \s_x \ = \ \vartheta \,
\s_t \ .
\end{eqnarray}
Note that for time-preserving symmetries the r.h.s. of both
equations vanish; we will thus deal with the slightly simpler -- and homogeneous -- equations
\begin{eqnarray}
\vphi_t & + & f \, \vphi_x \ - \ \vphi \, f_x \ + \ \frac12 \, \De
(\vphi ) \ = \ 0 \ , \label{eq:deteq1} \\
\vphi_w & + & \s \, \vphi_x \ - \ \vphi \, \s_x \ = \ 0 \ . \label{eq:deteq2}
\end{eqnarray}
Here and below, $\De$ is the \emph{Ito
Laplacian} \cite{\sderef}; in the scalar case this is given by
\beql{eq:Delta} \De (\phi) \ = \ \frac{\pa^2 \phi}{\pa w^2} \ + \
2 \, \s \ \frac{\pa^2 \phi}{\pa x \pa w} \ + \ \s^2 \ \frac{\pa^2
\phi}{\pa x^2} \ . \eeq

As well known, see e.g. \cite{Oksendal}, we can always reduce a nontrivial
(that is, with $\s \not= 0$) Ito equation \eqref{eq:Itogen} to the form with unit
noise coefficient ($\s = 1$) by the change of variable \beql{eq:xi} \xi \ = \
\int \frac{1}{\s (x,t)} \ d x \ . \eeq Thus, it will suffice to
consider the case with $\s (x,t) = 1$. The equation for $\xi(t)$ is
still an Ito equation, \beq d \xi \ = \ \Phi (\xi , t) \, dt \ + \ d w \ ; \eeq
the explicit expression of the new drift
coefficient in terms of the old drift and noise coefficients is
easily derived (see e.g. Sect.4 in \cite{GR22a}) and it turns out to be\footnote{In the r.h.s. of this, $x$ should be thought as a function of $\xi$ and $t$ through the change of variable inverse to \eqref{eq:xi}.}
\beq \Phi (\xi,t) \ = \  \frac{f}{\s} \ - \ \frac12 \, \s_x \ - \ \int \frac{\s_t}{\s^2} \, d x \ . \eeq

In the case of time-preserving symmetries, we know \cite{Koz1,Koz2,Koz3,Koz2018}
that an Ito equation \eqref{eq:Itogen} admitting a symmetry in the form
\eqref{eq:X} is integrated through the \emph{Kozlov substitution}, i.e.
passing to the dynamical variable \beql{eq:y} y \ = \ \int
\frac{1}{\vphi (x,t;w)} \ dx \ . \eeq The new equation for $y$ is
written in the form \beq dy \ = \ \Phi (t;w) \, dt \ + \ S (t;w) \, dw
\ ; \eeq when $\vphi_w = 0$ we also have $F_w = S_w = 0$ and hence
a proper Ito equation, while for $\vphi_w \not= 0$ in general we
get $F$ and $S$ also depending on $w$, hence a \emph{generalized
Ito equation} \cite{GR22a,GR22b,Gae23}.

\subsection{\FP equations}
\label{sec:gen:fp}

Let us now come to consider Fokker-Planck equations.
A general \FP equation for $u = u(x,t)$ is written in the
form
\beql{eq:FP0} \frac{\pa u}{\pa t} \ + \ \frac{\pa}{\pa x}
\[ \a(x,t) \ u\] \ - \ \frac12 \ \frac{\pa^2}{\pa x^2} \[ \b^2 (x,t) \
u \] \ = \ 0 \ . \eeq

Symmetry properties of \FP equations in one spatial dimension were
classified independently by Cicogna \& Vitali \cite{CV1,CV2} (focusing in particular on the time-autonomous case) and by Shtelen and Stognii \cite{ShS}; see also Spichak and Stognii \cite{SpS} and Stognii \cite{Sto} and the papers by Sastri and Dunn \cite{SD} and by Rudra \cite{Rudra}. The higher dimensional case was considered under additional assumptions by Spichak and Stognii \cite{SpS} and by Finkel \cite{Finkel}, and then more generally by Kozlov \cite{kozFP}. Symmetries of the \FP equation in connection with those of the underlying Ito equations were considered by Kozlov \cite{kozFP} (see also \cite{kozKB} for \emph{backward} Kolmogorov equations), but at the time only \emph{deterministic} symmetries were considered.

Here -- coherently with our approach focusing firstly and mostly on the simplest autonomous case -- we follow the simple discussion given in \cite{CV2}, which focuses on the autonomous case (see eq.(1) therein).\footnote{We provide a sketch of the computations leading to the Cicogna-Vitali classification in Appendix \ref{app:symmcv}, for convenience of the reader.}

First of all we note that there are some trivial symmetries,
and this independently of the form of functions $\a(x,t)$ and $\b(x,t)$. In
particular, we have (the vector field $Z_0$ being a symmetry only for the time-autonomous case) \beql{eq:trivsymm} Z_0 \ = \ \frac{\pa}{\pa t}
\ , \ \ Z_1 \ = \ u \, \frac{\pa}{\pa u} \ , \ \ Z_\zeta \ = \ \zeta \,
\frac{\pa}{\pa u} \ , \eeq where $\zeta$ is an arbitrary solution to
the \FP equation itself.

Here $Z_0$ is, as mentioned above, only present for autonomous equations (and is absent for general, time-dependent, \FP equations), while $Z_1$ and $Z_\a$ are also present in the general case; actually $Z_1$ corresponds to the fact the \FP equation is homogeneous (of degree one) in $u$, and $Z_\zeta$ reflects the (linear) superposition principle for solutions to linear homogeneous equations \cite{\symmref}.

In this sense the symmetries \eqref{eq:trivsymm} are trivial. We will not  consider such symmetries from now on, and only focus on nontrivial symmetries.

In the Cicogna-Vitali classification, one considers the function \beql{eq:gamma} \ga (x) \ := \ - \,
\frac12 \ \( \a^2 \ + \ \b^2 \, \a_x \)_x \ . \eeq We have then
different cases depending on properties of this function $\ga
(x)$.

\begin{itemize}

\item[(i)] If $\ga_{xx} = 0$, the \FP equation admits four
non-trivial symmetries $X_i$; their explicit expression (which is slightly involved)
can be determined and is given in \cite{CV2}, see eq. (9) therein (note these have a component along $\pa_t$); they are also reported in Appendix \ref{app:symmcv} here for ease of the reader. This means -- when one considers also the trivial symmetries listed in \eqref{eq:trivsymm} -- that in this case \emph{the symmetry algebra is isomorphic to that of the heat equation} \cite{Olver1}.\footnote{We note that in the case $\b = 1$ the equation $\ga_{xx} = 0$ can be explicitly solved for $\a$ in terms of hypergeometric functions and Hermite polynomials, and this both for the autonomous and the non-autonomous cases; thus this case can be explicitly characterized. Apparently this explicit characterization was not pursued by Cicogna and Vitali and by other authors. This is discussed (for the autonomous case, $\a = \a(x)$) in Section \ref{app:hypg} here.}

\item[(ii)] If $\ga_{xx} \not= 0$ and the equation \beql{eq:CV10}
\mathcal{G} \ := \ (\ga_x \ + \ \nu_1 ) \ (x \ + \ \nu_0 ) \ + \ 3
\, \ga \ = \ 0 \ , \eeq where $\nu_0,\nu_1$ are constants, admits a
solution then the \FP equation admits two nontrivial symmetries $X_i$; their explicit expression is given in \cite{CV2}, see eq. (12) therein, and again they are also reported in Appendix \ref{app:symmcv} here.

\item[(iii)] Finally, if $\ga_{xx} \not= 0$ but eq. \eqref{eq:CV10} admits no solution, then the \FP equation admits only the trivial symmetries
\eqref{eq:trivsymm}.

\end{itemize}
\par\noindent
Summarizing, the nontrivial symmetry algebra of an autonomous \FP equation can be of dimension zero, two, or four.
\medskip

Some remarks are in order about this conclusion:
\begin{enumerate}
\item In the first case (as already mentioned) the symmetry algebra is isomorphic to the symmetry algebra of the heat equation, which is the \FP equation corresponding to $\a (x,t)= 0$, $\b (x,t) = 1$.
\item When both the drift and the diffusion coefficients do not depend on $x$, and hence the Ito equation is immediately integrated, we have $\ga = 0$ and hence we are in the first case.
\item These conclusions may appear to be in some disagreement with what is obtained in \cite{GRQ1,GRQ2}; but one should note that these papers  considered, for what concerns symmetries of the \FP equation, only \emph{projectable} symmetries, while here we consider general ones.
\end{enumerate}

\subsection{Correspondence between an Ito and the associated \FP
equation}
\label{sec:gen:corr}

The correspondence between an Ito equation and the associated \FP
equation is very simple, even more so with the notation we are using.
Indeed, we just have that the coefficients $\a (x,t)$ and $\b(x,t)$
of the \FP equation \eqref{eq:FP0} associated to the Ito equation
\eqref{eq:Itogen} are given by  \beq \a(x,t) \ = \ f(x,t) \ , \ \
\b (x,t) \ = \ \s (x,t) \ . \eeq (We did not use this notation from
the beginning only to emphasize that the available results about
symmetries of \FP equations are completely independent of their
relation with stochastic equations.)

As we want to consider \FP equations associated to an Ito
equation, we will from now on write the \FP equation in the form
\beql{eq:FP} u_t \ + \ \pa_x \[ f(x,t) \, u \] \ - \ \frac12 \
\pa_x^2 \[ \s^2 (x,t) \, u \] \ = \ 0 \ . \eeq
As mentioned above, for a nontrivial Ito equation we can always,
by the change of variables \eqref{eq:xi}, reduce to $\s = 1$; correspondingly the
associated \FP equation will be of the form
\beql{eq:FPsimp} u_t \ + \ \( f_x \, u \ + \ f \, u_x \) \ - \ \frac12 \
  u_{xx} \ = \ 0 \ . \eeq

\section{Symmetries of scalar Ito equations}
\label{sec:symmito}

We now come to discuss symmetries of the Ito equations.

First of all, we note that once we have a general Ito equation
(and the associated \FP equation), we can always implement the
change of variable \eqref{eq:xi} and reduce to the case of
constant -- actually, unit -- noise. The new Ito equation will
have an associated \FP equation. As the change of variable is (by
definition) an isomorphism, the old and the new \FP equations are
also isomorphic, and in particular will have isomorphic symmetry
algebra, so we can just deal with the case of unit noise.

The classification of Ito equations of this type (that is,
possibly non-au\-to\-no\-mous) admitting symmetries obtained in
\cite{GR22a} is complete. We report here the result of that
discussion.

\medskip\noindent
{\bf Proposition 1.} \cite{GR22a} {\it A scalar Ito equation
with unit noise, \beq dx \ = \ f(x,t) \, dt \ + \ d w \ , \eeq
admits a time-preserving symmetry $X = \vphi (x,t,w) \pa_x$ if and
only if it corresponds to one of the following types:
\begin{itemize}
\item[(A)] $f(x,t) = h(t)$; in this case, denoting by $H(t)$ a
primitive of $h(t)$, the symmetry is identified by $$ \vphi = P
(\zeta) := P [x - w - H(t)] \ , $$ with $P$ an arbitrary smooth
function.
\item[(B)] $f(x,t) = h(t) + k(t) x$; in this case, denoting
by $K(t)$ a primitive of $k(t)$, the symmetry is identified by
$$ \vphi = \exp [ K(t)] \ . $$
\item[(C)] $f(x,t) = h(t) + k(t) \exp [\b
x]$, with $\b \not= 0$ a real constant; in this case, denoting by $H(t)$ a
primitive of $h(t)$, the symmetry is identified by $$ \vphi = \exp [
\b (x - w - H(t))] \ . $$
\end{itemize}}
\bigskip

This Proposition can be immediately restricted to the case of \emph{autonomous}
Ito equation; in this case we get (we stress that $h$ and $k$ are now real constants):

\medskip\noindent
{\bf Corollary 1.} \cite{GR22a} {\it A scalar \emph{autonomous} Ito equation
with unit noise, \beq dx \ = \ f(x) \, dt \ + \ d w \ , \eeq
admits a time-preserving symmetry $X = \vphi (x,t,w) \pa_x$ if and
only if it corresponds to one of the following types. with $h_0$ and $k_0$ real constants:
\begin{itemize}
\item[(A)] $f(x) = h_0$; in this case the symmetry is identified by $$ \vphi = P (\zeta) := P [x - w - h_0 t] \ , $$ with $P$ an arbitrary smooth
function.
\item[(B)] $f(x) = h_0 + k_0 x$; in this case the symmetry is identified by
$$ \vphi = \exp [ k_0 t] \ . $$
\item[(C)] $f(x) = h_0 + k_0 \exp [\b
x]$, with $\b \not= 0$ a real constant; in this case the symmetry is identified by $$ \vphi = \exp [ \b (x - w - h_0 t)] \ . $$
\end{itemize}}
\bigskip

It is worth pausing a moment to note that while for Ito equations
it makes perfect sense to consider random symmetries, i.e. maps of the dynamical variable $x$ depending on the realization of the $w(t)$ Wiener process, this makes little sense when we consider the associated \FP equation. In fact, the \FP description does not explicitly involve the driving Wiener process at
all. (Actually, it involves it in an implicit way, i.e. it describes an average over the realizations of the Wiener process; see the discussion later on in Sect.\ref{sec:sfp:disc}.)

This is a serious problem in view of the fact that in cases (A)
and (C) of the above classification, the symmetries do explicitly
depend on $w$. It should be noted that in case (A) we can somewhat
escape this problem by considering just a \emph{constant} function
$P (\zeta)$, in which case the dependence on $w$ disappears;
while this is not possible in case (C).

\section{Symmetries of the \FP equations associated to symmetric
Ito equations}
\label{sec:symmfp}

The prototypical (nontrivial) integrable Ito equation is of course $d x = d w$; its associated \FP equation is the heat equation $u_t = u_{xx}$, whose symmetry algebra is well known \cite{\symmref}; in the present context, this is a maximal symmetry algebra for equations of \FP type. A naive expectation could be that the integrable \FP equations corresponding to integrable Ito equations would have symmetry algebras isomorphic to that of the heat equation. We will see this is not the case.

In this Section, unless the contrary is explicitly stated, we will deal with \emph{autonomous} Ito and \FP equations.\footnote{We anticipate that this will identify a case in which the situation is different from what one could naively expect; we will then also consider the corresponding case for \emph{non-autonomous} equations.}

\subsection{Explicit computations}
\label{sec:sfp:comp}

As mentioned above, we can just consider the case where the noise coefficient is reduced to unity.
This means we are dealing with \FP equations of the form
\beql{eq:FPR} u_t \ + \ f \, u_x \ + \ f_x \, u \ - \ \frac12 \,
u_{xx} \ = \ 0 \ . \eeq Also, for $\s = 1$ the function $\ga$ defined in \eqref{eq:gamma} is given by \beql{eq:gammasimp}
\ga \ = \ - \, \frac12 \, \(f^2 \ + \ f_x \)_x \ = \ - \, \frac12
\ \( 2 \, f \, f_x \ + \ f_{xx} \) \ . \eeq

We now get promptly \beq \ga_{xx} \ = \ \frac{1}{2} \left(- \, 6 \,
f_x  \, f_{xx} \ - \ 2 \, f \, f_{xxx} \ - \ f_{xxxx} \right) \ . \eeq

We will now consider the cases of integrable scalar \emph{autonomous} Ito equations identified in our previous work \cite{GR22a}, see also Corollary 1 where they appear as cases (A), (B), and (C).

It is immediate to check that in cases (A) and (B) above we get
$\ga_{xx} = 0$, so we are in case (i) of the classification by
Cicogna and Vitali.

In fact, in case (A) we have $f(x,t) = h(t)$, and hence $\ga = 0$.
In case (B) we have $f(x,t) = h(t) + k(t) x$; hence by \eqref{eq:gamma} or \eqref{eq:gammasimp} we get
$ \ga =  -  k(t) [ h(t) + k(t) x ]$,
which of course entails $\ga_{xx} = 0$.

As for case (C), i.e.
\beq f(x,t) \ = \ h(t) \ + \ k(t) \ e^{\b x} \ , \eeq
some -- fully standard -- computations are
needed. We obtain \beq \ga \ = \ -\frac{1}{2} \, \beta \, e^{\beta
x} \, k(t)
   \left[ \beta \, + \, 2 \, h(t) \, + \, 2 \, k(t) \,  e^{\beta  x} \right] \ ;  \eeq
and from this
\beq \ga_{xx}   \ = \ - \, \frac12 \, \b^3 \, k(t) \, e^{\b \, x} \ \[ \b \ + \ 2 \, h(t) \ + \ 8 \, k(t) \, e^{\b x} \] \ . \eeq
This is nonzero, except for the (degenerate) case where $k(t) = 0$
or/and $\beta = 0$; note in this case we are back to case (A) above.

As $\ga_{xx} \not= 0$, we have to look for solutions to the equation \eqref{eq:CV10}. The function $\mathcal{G}$ now reads (omitting to indicate the $t$ dependence of $h$ and $k$ for ease of writing)
$$ \mathcal{G} \ = \ (\nu_0+x)
   \left(\nu_1 -\frac12 \b^2 k e^{\b x} (\b +2 h +4 k e^{\b x})\right)- \frac32 \b k e^{\b x}
    (\b + 2 h +2 k e^{\b x}) \ . $$
The equation $\mathcal{G}=0$ admits two solutions:
\begin{eqnarray*}
\nu_1 &=& 0 \ , \ \ \b \ = \ 0 \ , \\
\nu_1 &=& 0 \ , \ \ k(t) \ = \ 0 \ . \end{eqnarray*}
These are exactly the conditions under which we had $\ga=0$; as already observed, in this case we are then back to case (A).

In other words, we have obtained that the integrable Ito equation of case (C) does correspond to a \FP equation with \emph{trivial} symmetry algebra.

This same result is also obtained by means of explicit direct computation, thus independently of the Cicogna and Vitali classification; see Section  \ref{sec:directC}.

\subsection{Direct computation for case (C)}
\label{sec:directC}

We have seen that the nontrivial phenomenon, i.e. having an integrable Ito equation while the associated \FP equation has only trivial symmetries, arises in case (C) of our classification -- which falls within case (iii) of the Cicogna-Vitali classification, see Sect.\ref{sec:gen:fp}.

We want to show here that indeed the associated \FP equation has only trivial symmetries, not using the Cicogna-Vitali result but based on a direct explicit computation, thus confirming by this our result.

We consider would-be symmetry vector fields in the general functional form (see e.g. \cite{CV1,CV2} or \cite{kozFP} for the justification of such a restriction, which actually follows from the determining equations through explicit computations, used here to keep the length of the computation within reasonable limits)
\beql{eq:XC} X \ = \ \tau (t) \, \pa_t \ + \ \xi (x,t) \, \pa_x \ + \ \vphi (x,t,u) \, \pa_u \ . \eeq
We proceed then by the standard procedure, i.e. first compute the second prolongation $X^{(2)}$ of the vector field, then apply it to the \FP equation, and then restrict the result of this to the solutions of the \FP itself (this is implemented by substituting for $u_t$ according to the \FP equation). The result of this is an expression $\mathcal{R} (x,t,u,u_x,u_{xx})$, and we should require this to vanish identically.

As the dependencies on $u_x$ and $u_{xx}$ are explicit, we require the vanishing of the coefficients of any monomial in these variables; in particular we require the vanishing of the coefficients for $u_{xx}$ and $u_x^2$, and this yields
\begin{eqnarray}
\xi (x,t) &=& \chi (t) \ + \ \frac12 \, x \, \tau' (t) \ , \label{eq:xiC} \\
\vphi (x,t,u) &=& \phi_0 (x,t) \ + \ \phi_1 (x,t) \, u \ . \label{eq:phiC} \end{eqnarray}
We then specialize to the $f(x)$ corresponding to our case (C), i.e.
\beql{eq:fC} f(x) \ = \ h_0 \ + \ k_0 \ e^{\b x} \eeq and consider the coefficient of $u_x$. This determines the function $\phi_1 (x,t)$ as
\beq \phi_1 (x,t) \ = \ Q (t) \ + \ \frac12 \, \[ 2 \, k_0 \, e^{\b x} \, \chi (t) \ - \ 2 \, x \, \chi' (t) \ + \ \( h_0 + k_0 e^{\b x} \) \, x \, \tau' (t) \ - \ \frac12 \, x^2 \, \tau'' (t) \] \ . \eeq
At this point we look at the terms \emph{not} depending on $u$ in what remains of $\mathcal{R}$. These involve $\phi_0$, and more precisely their vanishing corresponds to
\beq \frac{\pa \phi_0}{\pa t} \ + \ \( h_0 + k_0 e^{\b x} \) \, \frac{\pa \phi_0}{\pa x} \ + \ \( k_0 \, \b \, e^{\b x} \) \, \phi_0 \ - \ \frac12 \, \frac{\pa^2 \phi_0}{\pa x^2} \ = \ 0 \ . \eeq
This is the requirement that $\phi_0$ satisfies the \FP equation (with the choice \eqref{eq:fC} for $f$), and we know this is always a solution in the case of linear equations.

At this point $\mathcal{R} = u \mathcal{R}_0$; the condition $\mathcal{R}_0 = 0$ provides a rather involved differential equation, which we do not report here. Taking its third $x$-derivative, we obtain
\begin{eqnarray*} \frac{\pa^3 \mathcal{R}_0}{\pa x^3} &=& \frac14 \, k_0 \, \b^3 \, e^{\b x} \, \[ 2 \, \b \, \( 2 h_0 + \b + 16 k_0 e^{\b x} \) \, \chi (t) \right. \\
& & \left. \ + \ \( 5 \b + 40 k_0 e^{\b x} + \b^2 x + 16 k_0 \b x e^{\b x} + 2 h_0 (5 + \b x) \) \tau' (t) \] \ . \end{eqnarray*}

Requiring this to vanish, we have
\beql{eq:appc} \frac{2 \, \b \, \( 2 h_0 + \b + 16 k_0 e^{\b x} \)}{5 \b + 40 k_0 e^{\b x} + \b^2 x + 16 k_0 \b x e^{\b x} + 2 h_0 (5 + \b x)} \ = \ - \ \frac{\tau' (t)}{\chi (t)} \ . \eeq

Here the l.h.s. is a function of $x$ alone, while the r.h.s. is a function of $t$ alone; thus each of these must be equal to the same constant $K$. Looking at the r.h.s. of \eqref{eq:appc}, this yields immediately
$$ \chi (t) \ = \ - \, K \, \tau' (t) \ . $$

When we look at the l.h.s. of \eqref{eq:appc}, things are more complex. We take the $x$ derivative of the l.h.s., and require this to vanish. This yields
\beq 16 \, \b^2 \, k \ \( 2 \, h_0 \ + \ \b \) \ e^{\b x} \ - \ 512 \, \( k_0 \, \b \)^2 \, e^{2 \b x} \ - \ 2 \, \b^2 \ \( 2 \, h_0 \ + \ \b \)^2 \ = \ 0 \ . \eeq
This vanishes only for $k_0 \b = 0$, i.e. when either $k=0$ or $\b = 0$ (or both). But in these cases, we are back to
$$ \begin{cases} f(x) = h_0 & (k_0 = 0), \\
f(x) = h_0 + k_0 x & (\b = 0). \end{cases} $$
These fall in cases (A) and (B), respectively, of our classification (see Proposition 1 in Sect.\ref{sec:symmito}).

This explicit computation does therefore confirm that in case (C) of our classification (for autonomous equations) the associated \FP equation admits only the trivial symmetries \eqref{eq:trivsymm}.

\subsection{Direct computation for the time-dependent case (C)}
\label{sec:directCTD}

In our discussion we have identified an interesting phenomenon, i.e. the fact that to a (formally) integrable Ito equation possessing a standard \emph{random} symmetry corresponds a \FP equation with \emph{trivial} symmetry algebra.

Our discussion was based on \emph{autonomous} scalar Ito equations, as these were sufficient to display the phenomenon mentioned above. On the other hand, as recalled in Section \ref{sec:symmito}, we also have a classification for \emph{time-dependent} integrable Ito equations, see Proposition 1. In this case too we have one case -- again case (C) -- in which integrability is due to the presence of a \emph{random} symmetry. Actually the functional form of the concerned integrable Ito equation is just the same as above, except that the constants $h_0$ and $k_0$ of eq.\eqref{eq:fC} are now replaced by smooth functions $h(t)$ and $k(t)$, see Proposition 1. Needless to say, adding a time-dependence is expected to just reduce the symmetry algebra of the equation, hence preserve the phenomenon of trivial symmetry algebra for the associated \FP equation; but here we check again that this is the case by means of a direct computation.

We consider the associated \FP equation, write down the determining equations, and separate the solutions corresponding to an arbitrary function $\zeta (x,t)$ satisfying the \FP equation itself for a vector field of the form \eqref{eq:XC}. We will refer to the remaining expression for the action of $X$ on the equation as $\mathcal{E}$; the coefficients of the various terms in it are the determining equations. The formulas \eqref{eq:xiC} and \eqref{eq:phiC} apply in this case as well. Next, instead of \eqref{eq:fC} we set
\beql{eq:fTC} f(x,t) \ = \ h(t) \ + \ k(t) \ \exp\[ \b \, x \] \ . \eeq
The equation corresponding to (the vanishing of) the coefficient of $u_x$, differentiated three times w.r.t. $x$, yields a third order equation for $\phi_1 (x,t)$, which can be explicitly solved. It results
\begin{eqnarray} \phi_1 (x,t) &=& q_0 (t) \ + \ q_1 (t) \, x \ + \ q_2 (t) \, x^2 \nonumber \\
& & \ + \ \frac12 \ e^{\b x} \, \( x \, k(t) \, \tau' (t) \ + \ \frac{2}{\b} \ \( k' (t) \, \tau (t) \ + \ \b \, k(t) \, \chi_0 (t) \) \) \ . \end{eqnarray}
This simplifies the coefficient of $u_x$; we then consider its derivative w.r.t. $x$: its vanishing corresponds to an equation which can now be solved, yielding
\beq q_2 (t) \ = \ - \, \frac14 \ \tau'' (t) \ . \eeq
Finally we consider the coefficient of $u_x$ itself; this yields -- when we require its vanishing -- an equation which is promptly solved, providing
\beq q_1 (t) \ = \ h' (t) \, \tau (t) \ + \ \frac12 \, h(t) \, \tau' (t) \ - \ \chi_0' (t)  \ . \eeq

At this point $\mathcal{E}$ does not contain any $u_x$ term any more; moreover now all the $x$ and $u$ dependencies are explicit, and we can thus consider the coefficients of different terms in these variables.

In particular, the coefficient of $x e^{2 \b x} u$ is made of only one term, i.e.
$$ \frac12 \, \b \, k^2 (t) \ \tau' (t) \ . $$ As both $\b$ and $k$ are assumed to be nonzero, this implies
\beq \tau (t) \ = \ c_1 \ . \eeq
Similarly, the coefficient of $e^{2 \b x} u$ yields
\beq \chi_0 (t) \ = \ - \, \frac{c_1}{\b} \ \frac{k'(t)}{k(t)} \ . \eeq

Looking now at the coefficient of $e^{\b x} u$, we have
\beq c_1 \ \( k(t) \, h' (t) \ - \ \frac{(k' (t))^2}{\b \, k(t)} \ + \ \frac{k'' (t)}{\b} \) \ = \ 0 \ . \eeq
This requires ({\bf case a}) \beql{appT:c1} c_1 \ = \ 0 \ , \eeq
unless ({\bf case b}) the functions $h(t)$ and $k(t)$ satisfy the relation
\beql{appT:h} h(t) \ = \ c_2 \ - \ \frac{k' (t)}{\b \, k(t)} \ . \eeq

\subsection*{Case (a)}

In case (a), i.e. for $c_1 = 0$, the full expression $\mathcal{E}$ is now reduced to $u q_0' (t)$, hence we get
\beql{appT:q0} q_0 (t) \ = \ c_2 \ . \eeq

In the end, we have (beside the symmetries $X_\zeta = Z_\zeta$ which we have not considered in this computation, see above), only one symmetry, i.e.
\beq X \ = \ u \pa_u \ = \ Z_1 \ . \eeq
Note that in this case the symmetry $Z_0$ appearing in \eqref{eq:trivsymm} is obviously absent, as we deal with a non-autonomous \FP equation.

\subsection*{Case (b)}

In case (b), i.e. for $c_1$ an arbitrary function and $h(t)$ satisfying \eqref{appT:h}, again the full expression $\mathcal{E}$ is now reduced to $u q_0' (t)$, and we get \eqref{appT:q0}. In this way we have got
\begin{eqnarray*}
\tau &=& c_1 \ , \\
\xi  &=& - \, \frac{c_1}{\b} \, \frac{k' (t)}{k (t)} \ , \\
\phi_1 &=& c_2 \, u \ . \end{eqnarray*}

In other words we have (beside the symmetries $X_\zeta = Z_\zeta $) two symmetries
\begin{eqnarray}
X_1 &=& \pa_t \ - \ \( \frac{k'(t)}{\b \, k(t)} \) \, \pa_x \ , \\
X_2 &=& u \, \pa_u \ . \end{eqnarray}
Note that the first one acts on the time variable; the second one is just the generator of a scaling symmetry $u \to \la u$, and corresponds to the linear character of the equation.

\subsection{Discussion}
\label{sec:sfp:disc}

The naive expectation that to an integrable Ito equation should correspond a \FP equation with maximal symmetry algebra is confirmed for integrable Ito equations of types (A) and (B) in the classification provided by Proposition 1. On the other hand, for integrable Ito equations of type (C) not only this expectation is not confirmed, but actually the associated \FP equation turns out to have a \emph{trivial} symmetry algebra.

There are two problems raised by this conclusion:
\begin{itemize}
\item[$(a)$] This seems to be in contradiction with the result, mentioned above, according to which symmetries of an Ito equation are reflected into symmetries of the corresponding \FP equation \cite{GRQ1};
\item[$(b)$] This is apparently contradicting the very idea of integrable equation.
\end{itemize}

\bigskip
\noindent
The situation needs some discussion.

\medskip\noindent
As for the point $(a)$, it should be noted that the result in \cite{GRQ1} concerned (in the language used in the present note) the correspondence between \emph{deterministic} symmetries of the Ito equation and (projectable) symmetries of the associated \FP equation. Or, type (C) integrable Ito equations have a \emph{random} symmetry; see Proposition 1. Random symmetries can not be reflected in symmetries of the \FP equation, as the latter does not involve any Wiener process. Thus this type of problem is not a real one.

\medskip\noindent
The situation is more delicate concerning the -- more substantial -- point $(b)$. In order to better understand the situation, it is convenient to consider how the symmetry integration takes place in this context, i.e. for random symmetries. (This point is discussed in some detail in a recent paper of ours \cite{Gae23}, so we will be rather quick, referring to that paper for details.)

Essentially, and along exactly the same path as in symmetry integration or reduction of deterministic ordinary differential equations \cite{\symmref}, once a symmetry \eqref{eq:X} of the Ito equation \eqref{eq:Itogen} has been determined, integration proceeds by a change of variable $(x,t,w) \to (y,t;w)$ with $y = y (x,t;w)$ such that in the new variables the symmetry vector field is written as $X = \pa_y$. This is a standard problem in the theory of characteristics \cite{ArnODE,ArnodeS}, and the solution is provided by the Kozlov substitution \eqref{eq:y}.

Now, the point is that if we are dealing with a \emph{deterministic} symmetry, i.e. if $\vphi = \vphi (x,t)$, this yields a proper change of coordinates in the $(x,t)$ space, i.e. $y = y(x,t)$, and we have a new Ito equation which is promptly integrated. To this integrable Ito equation corresponds a new \FP equation, in the form \eqref{eq:FPsimp}, which has maximal symmetry algebra (in fact, in the autonomous case for this it results $\ga = 0$, where $\ga$ is the function defined by Cicogna and Vitali).

But, in the case of \emph{random} symmetries, i.e. for $\vphi$ actually depending on $w$, the change of variables depends on the specific realization of the driving Wiener process.
In fact, in this case the Ito equation \eqref{eq:Itogen} is transformed into a \emph{generalized} Ito equation (see \cite{Gae23} for details), of the form
\beql{eq:genIto} dy \ = \ F (t;w) \, dt \ + \ S (t;w) \, d w \ . \eeq
As both $F$ and $S$ do not depend on the new dynamical variable $y$, this is integrable: for each realization of the driving process $w = w(t)$ we can perform integration and obtain
$$ y(t) \ = \ y(0) \ + \ \int_0^t F[s,w(s)] \, ds \ + \ \int_0^t S[s,w(s)] \, d w(s) \ . $$

On the other hand, the change of variable depends on the realization of the driving Wiener process, and indeed the transformed equation \eqref{eq:genIto} is not of proper Ito type, and hence has not an associated \FP equation.

In intuitive term, the \FP equation is providing an average of the evolution described by the Ito equation under different realizations of the Wiener process, and things become elementary for integrable Ito equations; but in this case the Ito equation corresponding to different realizations of the Wiener process are actually different equations, so their diffusion properties are also different.

\section{Characterization of drifts in the case (A)}
\label{app:hypg}

As mentioned in Section \ref{sec:gen:fp}, it is actually possible to fully classify the drifts falling in case $(i)$ of the Cicogna-Vitali classification \cite{CV1,CV2}, hence the (autonomous) \FP equations having a symmetry algebra isomorphic to that of the heat equation. In this Section, we show this and provide explicit formulas. Another phenomenon defeating naive expectations will show up as a byproduct of such a classification: we can have \FP equations with maximal symmetry while the associated Ito equation has no symmetry.

We consider $\ga$ as defined by \eqref{eq:gamma}, and set $\s = 1$; as discussed above, this is always possible via the change of variable \eqref{eq:xi}. The \FP equations we are considering are hence of the form \eqref{eq:FPsimp}.

The condition for this equation to have a symmetry group isomorphic to that of the heat equation can be written \cite{CV2} as $\ga_{xx} = 0$. In view of the definition \eqref{eq:gamma} of $\ga$, this corresponds to a differential equation to be satisfied by the drift $f=f(x)$, i.e.
\beq\label{Bfourth}
\left( f^2 \ + \  f'\right)_{xxx}\ = \ 0
\eeq
This is a fourth order nonlinear ordinary differential equation, written explicitly as\footnote{We mention that feeding this into {\tt Mathematica}, or a similar symbolic manipulation program, one obtains an explicit solution; the expression of this, however, is quite complex and not illuminating. We will see that one instead can make contact with classical (special) functions.}
\beq
f^{(4)} \ + \ 2 \, f \, f''' \ + \ 6 \, f' \, f'' \ = \ 0 \ .
\eeq

However, due to its particular form it is clear that \eqref{Bfourth}  can be written as a first order differential equation:
\begin{eqnarray}
f' \ + \ f^2 &=& p(x) \ , \label{Bfirst}
 \\
p(x) &:=& \mu_0 \ + \ \mu_1 \, x \ + \ \mu_2 \, x^2 \nonumber \end{eqnarray}
where $\mu_i$ ($i=0,1,2$) are arbitrary constants.

This is not the most appropriate form to study the equation according to the classical theory of special functions. We thus define a new function $u(x) > 0$ via
\beq
f(x) \ = \ \frac{d}{dx} \log u(x) \ = \ \frac{u'(x)}{u(x)}
\eeq
and substitute into the equation \eqref{Bfirst}, getting
\beq\label{BfirstU}
\frac{u''}{u} \ - \ \frac{u'^2}{u^2} \ + \ \frac{u'^2}{u^2} \ = \ p(x) \ ;
\eeq
that is,
\beq\label{weber}
 u''(x) \ - \ p(x)\, u(x) \ = \ 0 \ .
\eeq
This is a second order linear equation, known as the \emph{Weber equation}; its solutions are the \emph{parabolic cylinder functions}\footnote{Properties of parabolic cylinder functions and details about them are given e.g. in  {\tt https://dlmf.nist.gov/12}.} and denoted as $D$.

The standard Weber equation (originated in the separation of the Laplace equation in parabolic cylinder coordinates), providing the  parabolic cylinder  functions, is written as
\beq\label{Bstandard}
v''(z)\ + \ \( \lambda \ + \ \frac12 \ - \ \frac14 \, z^2 \) \ v(z) \ = \ 0
\eeq
with general solution
\beq
v(z) \ = \ c_1 \ D(\lambda,z) \ + \ c_2 \, D(-1-\lambda,\mathrm{i}\, z)
\eeq

The parabolic cylinder functions are related to Hermite polynomials when the parameter $\lambda$ is an integer $n\ge 0$,
\beq
D(n,z)\ = \ 2^{-n/2}\, e^{-z^2/4}\,H_n \( z / \sqrt{2} \) \ .
\eeq

Let us construct a change of variables in equation \eqref{weber} leading to the classical Weber equation \eqref{Bstandard}. The second degree polynomial $p(x)$ can be written (in a generic case) as
\begin{eqnarray}
p(x) &=& \mu_2\, x^2 \ + \ \mu_1\,x \ + \ \mu_0 \nonumber \\
&=&  \mu_0 \ -\ \frac{\mu_1^2}{4\,\mu_2}\ + \  \left(\sqrt{\mu}_2\,x\ +\ \frac{\mu_1}{2\,\sqrt{\mu_2}}\right)^2 \ , \end{eqnarray}
and the equation \eqref{weber} is then
\beq
 u'' \ + \ \left[\frac{\mu_1^2}{4\,\mu_2}\ - \ \mu_0 \ - \  \left(\sqrt{\mu}_2\,x\ +\ \frac{\mu_1}{2\,\sqrt{\mu_2}}\right)^2\right]\, u(x) \ = \ 0 \ .
\eeq
Let us introduce the new variable $z$
\beq
z \ := \  \left(\frac{4}{\mu_2}\right)^{1/4}\,\left(\sqrt{\mu_2}\,x\ +\ \frac{\mu_1}{2\,\sqrt{ \mu_2}}\right) \ , \eeq so that \beq \frac{d^2u}{d x^2}\ =\  2\,\sqrt{\mu_2}\, \frac{d^2 u}{d  z^2} \ .
\eeq
The new equation is
\beq
  2\,\sqrt{\mu_2}\,u'' \ + \ \left[\frac{\mu_1^2}{4\,\mu_2 } \ - \   \mu_0   \ - \  \frac{\sqrt{\mu_2
}}{2}\,z^2 \right]\, u(x) \ = \ 0 \ ;
\eeq
this can be rearranged to give
\beql{eq:66}
 u'' \ + \ \left[\frac{\mu_1^2}{8\,\mu_2^{3/2}}\ - \   \frac{\mu_0}{2\sqrt{\mu_2}}   \ - \  \frac{z^2}{4}\ \right]\, u(x) \ = \ 0 \ .
\eeq
We define the parameter $\lambda$ as
\beq
\lambda \ = \ \frac{\mu_1^2}{8\,\mu_2^{3/2}} \ - \   \frac{\mu_0}{2\sqrt{\mu_2}}   \ - \ \frac12 \ .
\eeq
Then the drift $f(x)$ can be written as solutions to this equation \eqref{eq:66}, hence in terms of parabolic cylinder functions.

\subsection{A simple example}

Since parabolic cylinder functions are related to Hermite polynomials for certain values of the parameters, let us construct a particular solution in this case. Take $\mu_0=0$, $\mu_1=2\sqrt{3}$ and $\mu_2=1$.
Then,
\beq
\lambda=1
\eeq
and a solution of the Weber equation is:
\beq
u(z)\ = \ D(1,z)\ = \ 2^{-1/2}\, e^{-z^2/4}\,H_1(z/\sqrt{2})\ = \ z\, e^{-z^2/4}
\eeq
The original variable $x$ is:
\beq
z \ = \  \sqrt{2}\, x\ +\ \sqrt{6}
\eeq
and then,
\beq
u(x)=(\sqrt{2}\, x\ +\ \sqrt{6})\, e^{-x^2/2 \, +\, \sqrt{3}\,x\, -\, 3/2}
\eeq
Finally
\beq
f(x)\ = \ \frac{u'(x)}{u(x)}\ =\ \frac{d}{dx}\log u(x)\ =\ \frac{1}{x+\sqrt{3}} \ - \ \( x + \sqrt{3}\)
\eeq
It is straightforward to check that this drift satisfies $\gamma_{xx}=0$, and then the corresponding \FP equation is equivalent to the heat equation and maximally integrable (with a parametric group of six parameters).

As for the Ito equation, with this drift we have
\beq
dx \ = \  \left(\frac{1 }{x+\sqrt{3}}\ -\ x-\sqrt{3}\right)\, dt + \ dw \ .
\eeq
This Ito equation does not admit any symmetry, as follows from the classification in \cite{GR22a}, and as can also be checked by direct computation with the determining equations \eqref{eq:deteq1}, \eqref{eq:deteq2}. (The direct computation yields the same result -- that is, no symmetry -- also considering the determining equations for W-symmetries.  Similarly, a direct computation can be performed to check this \FP equation has indeed maximal symmetry.)

Thus, this example provides a counterpart to the situation examined in Sections \ref{sec:directC} and \ref{sec:directCTD}, and discussed in some detail  in Section \ref{sec:sfp:disc}. In that case, we had a symmetric Ito equation (more precisely, a W-symmetric one) whose associated \FP equation admitted no symmetry. Here we have met a \FP equation with maximal symmetry whose associated Ito equation admits no symmetry.

\section{Conclusions}
\label{sec:conclu}

We have considered the correspondence between symmetries of integrable scalar (autonomous) Ito equations and those of the associated \FP equations. We have found that in the case of Ito equations which possess standard \emph{deterministic} symmetries -- and hence are integrated via the Kozlov procedure by passing to variables adapted to such symmetries -- the associated \FP equations possess a maximal symmetry algebra, i.e. an algebra isomorphic to that of the heat equation, which in this context should be seen as the \FP equation associated to the free Ito equation $dx = dw$. On the other hand, for Ito equations which possess standard \emph{random} symmetries -- and hence are \emph{formally} integrated via the Kozlov procedure by passing to variables adapted to such symmetries -- the associated \FP equations possess only a trivial symmetry algebra, i.e. an algebra corresponding to ``generic'' \FP equations.

Having identified the ``interesting'' phenomenon in the autonomous case, we then passed to consider the general, non-autonomous, case; we have checked, by explicit computations, that the phenomenon we identified is also present in this  general time-dependent case.

In Section \ref{app:hypg} -- while providing a full description of autonomous \FP equations with maximal symmetry in terms of classical special functions (which appears to be new) -- we have seen that this ``interesting phenomenon''  has a counterpart with roughly inverted roles between the Ito and the \FP equations: we can have a \FP equation with \emph{maximal} symmetry algebra and the associated Ito equation having \emph{no} symmetry (both standard symmetry and  W-symmetry) at all.

Our work confirms what was stated by Kozlov (dealing with a different set of admitted transformations). In the conclusions of his work \cite{kozFP}, he writes: {\it ``using symmetries of stochastic differential equations we can obtain only \emph{partial} results for symmetries of the \FP equation''}.

We have shown that this is even more true when one considers also random symmetries. In fact -- contrary to the naive expectations one could have -- it turned out that to an Ito equation which is integrable thanks to a \emph{random} symmetry may correspond a \FP equation which has only a trivial symmetry.

Finally we note that this conclusion calls for an intrinsic characterization of \FP equations corresponding to Ito equations which are (formally) integrable via standard random symmetries. Such a characterization can of course be obtained by just listing the \FP equations corresponding to the Ito equations with the property mentioned above -- which were classified in our recent work \cite{GR22a}; but one wonders what these \FP equations have in common in intrinsic terms, i.e. without looking at the associated Ito equations.

\section*{Acknowledgements}

\small This paper grew out of a question posed by Giorgio Gubbiotti, whom we warmly thank. MAR acknowledges the  support of Universidad Complutense de Madrid under grant G/6400100/3000. The paper was written while GG was on sabbatical leave from Universit\`a di Milano; in this time he was in residence at, and enjoying the relaxed atmosphere of, SMRI. The work of GG is also supported by GNFM-INdAM.

\begin{appendix}

\section{Nontrivial symmetries of the autonomous \\ Fokker-Planck equation}
\label{app:symmcv}

In this Appendix we sketchily review, for convenience of the reader, the computations leading to the Cicogna-Vitali classification \cite{CV1,CV2} of the symmetries of autonomous scalar \FP equation we have been using -- but of course adapting their notation to the one we are using here. We will just consider the case where $\s = 1$, as one can always reduce to this case in a standard and simple way, see \eqref{eq:xi}. That is, we consider symmetries of \eqref{eq:FPsimp}.

Consider a vector field:
\beq
X \ = \ \tau(t,x,u)\,\pa_t\  + \ \xi(t,x,u)\,\pa_x\  + \ \phi(t,x,u)\,\pa_u
\eeq
In order that $X$ be a symmetry of the \FP equation, their coefficients must satisfy the determining equations which are obtained from its second order prolongation, when applied to  the \FP equation \cite{Olver1}:
\beq
\mathbf{pr}^{(2)}X=X+\phi^x\pa_{u_x}+\phi^t\pa_{u_t}+\phi^{xx}\pa_{u_{xx}}=0
\eeq

Once the \FP equation has been used to eliminate the term $u_{xx}$, the determining equations reduce the dependence of the unknown  functions $\xi,\tau,\phi$ on the variables $x,t,u$; more precisely, we get
\beq
\begin{cases}
\tau(t,x,u)\  = \ \tau(t) \ , \\
\xi(t,x,u) \  = \ \xi(t,x) \ , \\
\phi(t,x,u)\  = \ \phi_0(t,x) \ + \ \phi_1(t,x) \, u \ .
\end{cases}
\eeq
Using this result we get a set of four differential equations (since the dependence on $u$ is explicit). The first one is:
\beq\label{first}
\xi_x \ = \  \frac12 \, \tau_t \ ,
\eeq
which can be easily integrated (up to now, $\chi(t)$ is an arbitrary function of $t$) as
\beq
\xi(x,t) \ = \   \frac12 \, x \, \tau_t(t) \ + \ \chi(t) \ .
\eeq
The second equation is
\beq
\phi_{0,t} \ - \ \frac12 \, \phi_{0,xx} \ + \ f_x \, \phi_0 \ + \ f\, \phi_{0,x}  \  = \ 0 \ ;
\eeq
this expresses the fact that $\phi_0(x,t)$ is a solution of the equation. The two other equations can be written as:
\begin{eqnarray}
 \phi_{1,x}&=&  \frac{\pa}{\pa x}\left(  - \ x\,\chi_t\ + \ \chi\,  f  \ +\ \frac12\, x \, f   \,\tau_t \  -\ \frac14\,x^2 \,\tau_{tt}\right) \ , \label{fir}\\
\phi_{1,t} &=&   \ -\  \chi\, f_{xx}\ - \ \frac{1}{2}\, (x\,f)_{xx} \,\tau_t  \ + \ \frac{1}{2}\, \phi_{1,xx}\ -\  \phi_{1,x}\, f \ . \label{sec}
\end{eqnarray}
Equation \eqref{fir} can be easily integrated, yielding
\beql{eq:82}
 \phi_1(x,t)\ =\    - \ x\,\chi_t(t)\ + \ \chi(t)\,  f(x)  \ +\ \frac12\, x \, f(x)  \,\tau_t(t) \  -\ \frac14\,x^2 \,\tau_{tt}(t)\ + \ g(t)
 \eeq
where $g(t)$ is an arbitrary function of $t$. Substituting in \eqref{sec} and defining \cite{CV2}
\beql{eq:gammapp}
\gamma(x) \ = \ - \ \frac12 \, \( f(x)^2 \ + \  f_{x}(x) \)_x \ ,
\eeq
we get
\beq
x \, \chi_{tt} \ + \ \gamma \, \chi \ = \ - \frac12 \( \frac{1}{2} \, x^2 \, \tau_{ttt} \ - \ \(  - x \, \gamma \ + \ f_x \ + \ f^2 \) \tau_t \) \ + \ \frac{\tau_{tt}}{4} \ + \ g_t \ .
\eeq
This is an equation mixing functions of $x$ and $t$ (and their derivatives).
If we take successive derivatives in $x$, we get the equations:
\begin{eqnarray}
 \chi_{tt}\ +\ \gamma_x\, \chi &=& - \ \frac{1}{2}\,\left( x\, \tau_{ttt} \ +\  (3\, \gamma\ + \ x\,\gamma_x) \,\tau_t\right) \ , \label{der1} \\
 \gamma_{xx}\, \chi &=& -\ \frac{1}{2}\,\left( \tau_{ttt} \ +\  (4\,\gamma_x \ + \  \,x\,\gamma_{xx})\,\tau_t\right) \ , \label{der2} \\
\gamma_{xxx}\, \chi &=& -\ \frac{1}{2}\,\left(   5\,\gamma_{xx} \ +\ x\,\gamma_{xxx} \right) \,\tau_t \ . \label{der3} \end{eqnarray}

We will distinguish the following cases.

\subsection{Case (i)}

Assume $\gamma_{xx}\ =\ 0$;
that is,
\beq
\gamma(x) \ = \ - \, \mu_2 \, x \ - \ \frac12 \, \mu_1 \ .
\eeq
Then
\beq\label{firstorder1}
f'(x) \ = \  \mu_0 \ + \ \mu_1 \, x \ + \ \mu_2 \, x^2 \ - \ f(x)^2 \ ,
\eeq
where $\mu_i$, $i=0,1,2$ are arbitrary constants (we will not need the explicit form of $f(x)$, which  in the general case is rather complicated). We get an equation for $\tau$
\beq\label{tau1}
\tau_{ttt} \ - \  4 \, \mu_2 \, \tau_t \ = \ 0 \ .
\eeq
Substituting into \eqref{der1} we have
\beql{der4}
 \chi_{tt} \ - \ \mu_2 \, \chi \ = \ \frac{3}{4} \, \mu_1 \, \tau_t \ ,
\eeq
and equations \eqref{tau1} and \eqref{der4} allow to compute $h(t)$ and $\tau(t)$ easily.

As discussed in \cite{CV2} the condition $\gamma_{xx}=0$ allows to transform the \FP equation into the heat equation, and then, in this case, the symmetry group of the \FP equation is isomorphic to that of the heat equation. Let us compute the symmetry vector fields for the \FP equation. Solving equation \eqref{tau1} we have
\beq
\tau (t) \ = \ \lambda_1 \, e^{2 \sqrt{\mu_2}\, t} \ + \ \lambda_2 \, e^{-2 \sqrt{ \mu_2}\, t} \ + \ \lambda_3 \ .
\eeq
We then compute $\chi(t)$ and finally $g(t)$. The nontrivial symmetry vector fields are
\begin{eqnarray}
X_1 &=& \frac{e^{2 \sqrt{\mu_2} t}}{2 \sqrt{\mu_2}} \, \pa_t \ + \ \frac12 \, \(x + \frac{\mu_1}{2 \mu_2}\) \, e^{2 \sqrt{\mu_2} t} \, \pa_x \nonumber \\
& & \ + \ \frac12 \,  e^{2 \sqrt{\mu_2}t}  \, \( f(x) \, \(x + \frac{\mu_1}{2 \mu_2} \) \ - \ \zeta  \ - \ \frac{1}{2} \) \, u \, \pa_u \ , \nonumber \\
X_2 &=& - \, \frac{e^{- 2 \sqrt{\mu_2} t}}{2 \sqrt{\mu_2}} \, \pa_t \ + \ \frac12 \, \(x + \frac{\mu_1}{2 \mu_2}\) \, e^{- 2 \sqrt{\mu_2} t} \, \pa_x \nonumber \\
& & \ + \ \frac12 \,  e^{- 2 \sqrt{\mu_2}t}  \, \( f(x) \, \(x + \frac{\mu_1}{2 \mu_2} \) \ + \ \zeta  \ - \ \frac{1}{2} \) \, u \, \pa_u \ , \nonumber \\
X_3 &=&  \, e^{\sqrt{\mu_2} t} \, \pa_x \ + \  e^{\sqrt{\mu_2} t}  \, \( f(x) \ - \ \sqrt{\mu_2} \, \( \frac{\mu_1}{2 \mu_2} + x \) \) \, u \, \pa_u \ , \nonumber \\
X_4 &=&   \, e^{- \sqrt{\mu_2} t} \, \pa_x \ + \  e^{- \sqrt{\mu_2} t}  \, \( f(x) \ + \ \sqrt{\mu_2} \, \( \frac{\mu_1}{2 \mu_2} + x \) \) \, u \, \pa_u \ .
\end{eqnarray}
Here we have written
\beq \zeta  \ := \ \sqrt{\mu_2} \, x^2 \ + \ \frac{\mu_1}{\sqrt{\mu_2}} \, x \ + \ \frac{\mu_0}{2 \sqrt{\mu_2}} \ + \ \frac{\mu_1^2}{8 \mu_2 \sqrt{\mu_2}} \ . \eeq
Note that $X_1$ and $X_2$ have a component along $\pa_t$, while $X_3$ and $X_4$ do not.

\subsection{Case (ii)}

Let us now consider the case where $\ga_{xx} \not= 0$ and $\gamma_{xxx} \not= 0$ as well. Now we get
\beq
\chi \ = \ -\ \left(\frac{5\,\gamma_{xx}}{2\,\gamma_{xxx}}\ +\ \frac{x}{2 }\right)\,\tau_t \ ,
\eeq
which implies
\beq
\frac{5\,\gamma_{xx}}{2\,\gamma_{xxx}}\ +\ \frac{x}{2 }\ =\ -\frac12\,\nu_0 \ ;
\eeq
this has solution  \cite{CV2}
\beq
\gamma(x)\ = \  c\ -\ \frac{b}{ (x\,+\,\nu_0)^3}\ -\ \frac{\nu_1}{4}\, x \ .
\eeq
This equation yields a special form of the drift, which satisfies the equation
\beq
f'(x)\ = \  -\ f(x)^2\ -\ \frac{b}{( \nu_0\ +\ x)^2}\ -\ 2\, c\, x\ +\ \frac{ \nu_1 }{4}\,x^2\ +\ \zeta \ .
\eeq

The equations for $\tau(t)$ and $\chi(t)$ are:
\beq
\tau''' \ - \ \nu_1 \, \tau' \ = \ 0 \ , \ \ \  \chi(t) \ = \ \frac12 \, \nu_0 \, \tau' \ .
\eeq
Solving the equation for $g(t)$ appearing in \eqref{eq:82}, we find the constraint:
\beq
c \ = \ - \, \frac14 \, \nu_0 \, \nu_1
\eeq
and this yields, after a straightforward computation, the vector fields
\begin{eqnarray*}
X_1 &=& \frac{e^{\sqrt{\nu_1} t}}{\sqrt{\nu_1}} \, \pa_t \ + \ \frac12 \, (x + \nu_0) \, e^{\sqrt{\nu_1} t} \, \pa_x \nonumber \\
& & \ + \ \frac12 \,  e^{\sqrt{\nu_1}t}  \, \( f(x) \, (x + \nu_0) \ - \ \frac{\sqrt{\nu_1}}{2} \, (x + \nu_0)^2 \ + \ \frac{\rho}{\sqrt{\nu_1}} \ - \ \frac{1}{2} \) \, u \, \pa_u \ , \nonumber \\
X_2 &=& - \, \frac{e^{- \sqrt{\nu_1} t}}{\sqrt{\nu_1}} \, \pa_t \ + \ \frac12 \, (x + \nu_0) \, e^{- \sqrt{\nu_1} t} \, \pa_x \nonumber \\
& & \ + \ \frac12 \, e^{- \sqrt{\nu_1}t}  \, \( f(x) \, (x + \nu_0) \ + \ \frac{\sqrt{\nu_1}}{2} \, (x + \nu_0)^2 \ - \ \frac{\rho}{\sqrt{\nu_1}} \ - \ \frac{1}{2} \) \, u \, \pa_u \ .\end{eqnarray*}
Here we have written:
$$ \rho \ := \ \frac{ \nu_0^2 \,\nu_1}{4}\ -\ \zeta \ . $$

\subsection{Case (iii)}

Finally, consider the case where $\gamma_{xx} \not= 0$ with $\gamma_{xxx}=0$.
Now eq. \eqref{der3} reads
\beq\label{der3b}
\gamma_{xxx}\, \chi \ = \ -\ \frac{1}{2}\,\left(   5\,\gamma_{xx} \ +\ x\,\gamma_{xxx} \right) \, \tau_t \ ; \ \ \
\gamma_{xx}  \,\tau_t \ = \ 0 \ .
\eeq
Then $\tau(t) \, = \, \kappa$, a constant, and $\chi(t) \, = \, 0$  \cite{CV2}. In this case the symmetry vector field is
\beq
X \ = \ \kappa \, \pa_t \ + \ \( \lambda \, u \ + \ \mu(x,t) \) \, \pa_u
\eeq
where $\kappa$, $\lambda$ are arbitrary constants and $\mu(x,t)$ is a solution of the \FP equation itself; that is, we get only trivial symmetries.

\end{appendix}

\label{lastpage}
\end{document}